\documentclass[prl, amsmath,twocolumn,longbibliography]{revtex4-1}
\usepackage{graphicx}
\usepackage{color}
\usepackage{latexsym}
\usepackage{amssymb}

\newcommand{\eps}{\varepsilon}

\newcommand\cJ{{\mathcal J}}
\newcommand\cG{{\mathcal G}}

\begin{document}

\title{Phase transition in protocols minimizing work fluctuations}

\author{Alexandre P. Solon}
\author{Jordan M. Horowitz}
\affiliation{Physics of Living Systems Group, Department of Physics, Massachusetts Institute of Technology, 400 Technology Square, Cambridge, Massachusetts 02139}

\date{\today}

\begin{abstract}
  For two canonical examples of driven mesoscopic systems -- a
  harmonically-trapped Brownian particle and a quantum dot -- we
  numerically determine the finite-time protocols that optimize the
  compromise between the standard deviation and the mean of the
  dissipated work. In the case of the oscillator, we observe a
  collection of protocols that smoothly trade-off between average work
  and its fluctuations. However, for the quantum dot, we find that as
  we shift the weight of our optimization objective from average work
  to work standard deviation, there is an analog of a first-order
  phase transition in protocol space: two distinct protocols exchange
  global optimality with mixed protocols akin to phase coexistence.
  As a result, the two types of protocols possess qualitatively
  different properties and remain distinct even in the infinite
  duration limit: optimal-work-fluctuation protocols never coalesce
  with the minimal work protocols, which therefore never become
  quasistatic.
\end{abstract}

\maketitle 
 
Essential to any well-functioning thermodynamic engine is the rapid
and reliable extraction of work at high thermodynamic efficiency.
Accomplishing this goal requires both characterizing the optimal
finite-time protocols that maximize the work extracted (or minimize
the work dissipated)~\cite{Schmiedl2007, Then2008, Esposito2010,
  Aurell2011} and understanding the trade-off (or lack thereof) with
the engine
efficiency~\cite{Schmiedl2008,Chvosta2010,Moreau2012,Allahverdyan2013,Holubec2014,Shiraishi2016,Raz2016,Proesmans2016}.
Arguably though, large power with high efficiency is only useful when
the cycle-to-cycle fluctuations are small.  Thus, it is equally as
important to characterize any trade-offs with power
fluctuations~\cite{Ritort2004, Holubec2014,Funo2015,Holubec2017}.

One place where universal statements about power fluctuations can be
made is in autonomous thermodynamic heat engines -- those driven by a
constant flow of heat down a temperature gradient.  For these
stationary engines, the thermodynamic uncertainty
relation~\cite{Barato2015,Pietzonka2016Universal,Pietzonka2017,Pietzonka2016Affinity,Gingrich2016Bound,Gingrich2017,Gingrich2017FPT,Horowitz2017,Dechant2017,Nardini2017}
imposes a universal trade-off between power, power fluctuations, and
thermodynamic efficiency~\cite{Pietzonka2017universal}. One
might hope that such a universal trade-off exists for nonautonomous
thermodynamic engines -- driven by cyclic variations of an external
parameter.  Counterexamples, however, invalidate the naive
extension of this
prediction~\cite{Ritort2004,Barato2016,Piglotti2017b,Proesmans2017,Chiuchiu2017}.

\begin{figure}
\includegraphics[height=3.4cm,width=\columnwidth]{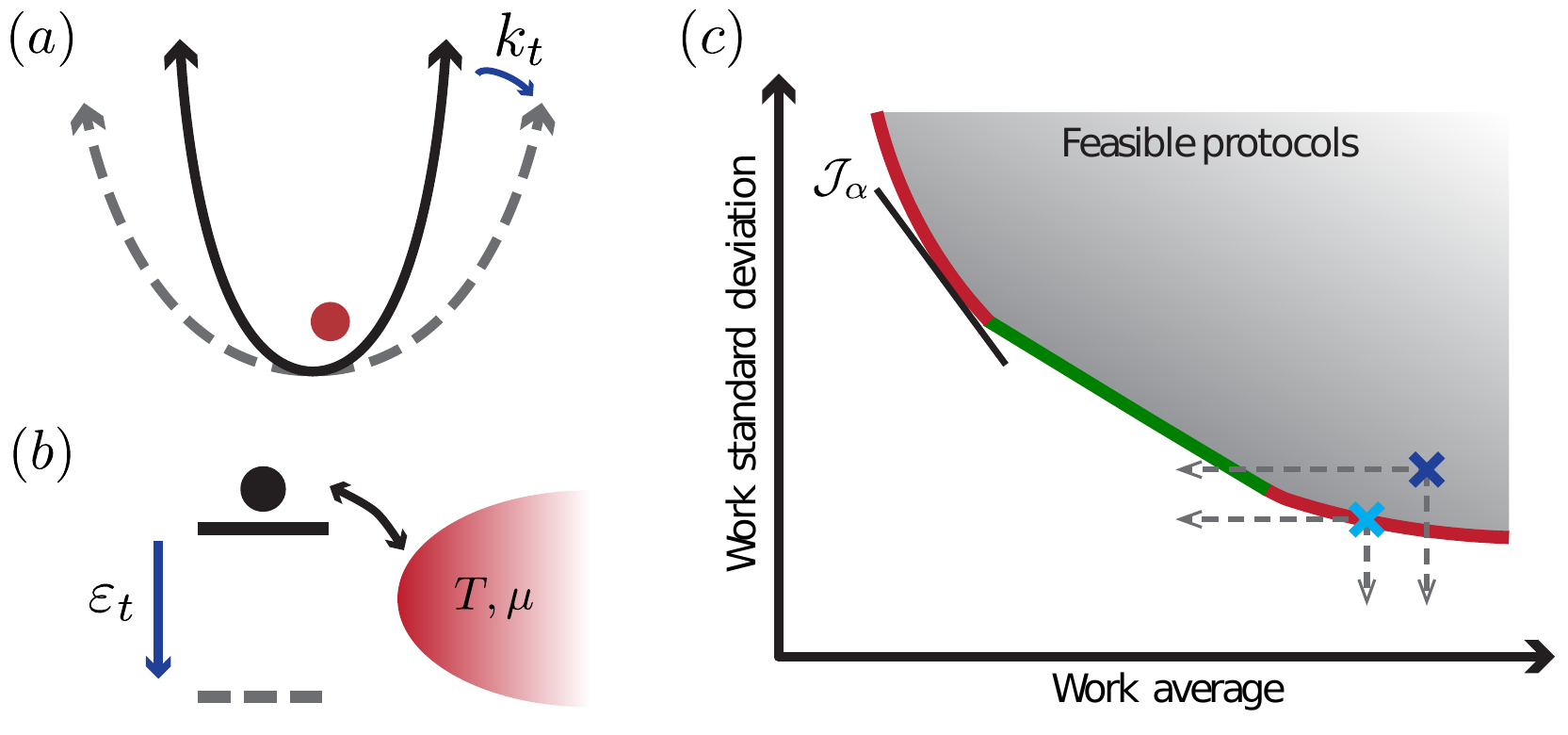}
\caption{{\bf (a):} Harmonically-trapped Brownian particle with an
  expanding spring constant $k_t$. {\bf (b):} Quantum dot exchanging
  particles with a reservoir at temperature $T$ and chemical potential
  $\mu$ with a decreasing energy $\varepsilon_t$.  {\bf (c):} Pareto
  front (red/green) bounds the region of allowed
  protocols. Sub-optimal protocol (dark blue cross) is dominated by
  all protocols down and left, including the nondominated protocol
  (light blue cross).  The single objective ${\mathcal J}_\alpha$,
  represented by a black line for fixed $\alpha$, which is optimal
  when tangent to the front.  All solutions along the flat green
  portion correspond to the same $\alpha$.}
  \label{fig:illustration}
\end{figure}

It thus remains to characterize optimal power fluctuations in driven
nonautonomous engines.  As a first step, we
investigate finite-time thermodynamic processes that attempt to
minimize both the work fluctuations and the average dissipated work.
Specifically, for two canonical models of driven mesoscopic systems --
a harmonically-trapped Brownian particle~\cite{Schmiedl2007} and a
quantum dot~\cite{Esposito2010}, illustrated in
Fig.~\ref{fig:illustration} -- we numerically determine the collection
of protocols that optimize the compromise between the average and
standard deviation of the work.  Remarkably, for the quantum dot, as
we shift the weight of our optimization objective from average work to
work standard deviation, we observe the analog of a first-order phase
transition, featuring two distinct local minima in protocol space that
exchange global optimality and mixed protocols akin to phase
coexistence. Looking at protocols of increasing duration, we show
that protocols minimizing work fluctuations need not be quasistatic in
the infinite time limit, and thus remain out of reach of a linear
theory.

Lastly, we adopt here the standard deviation as our metric for the
magnitude of work fluctuations, largely because it naturally appears
in the thermodynamic uncertainty relation, and near equilibrium has
universal
properties~\cite{Hermans1991,Jarzynski1997,Speck2004,Piglotti2017b}.
However, there are other ways to characterize work
fluctuations. Integrated squared power lends
itself to analytic treatment using optimal control
theory~\cite{Filliger2005}. Alternatively, single-shot thermodynamics
has emerged as a program that allows one to design protocols that make
very large fluctuations extremely
unlikely~\cite{Aberg2013,Halpern2015}. Finally, the authors of
Ref.~\cite{Ziao2014} numerically optimized the exponential average of
the work.

{\it Setup.---} We have in mind a mesoscopic system with states $x$ --
continuous or discrete -- evolving in a noisy thermal environment at
temperature $T$, under the influence of an externally-controlled
potential $U(x,\lambda)$. The system is driven by a protocol
$\lambda_t$ during a finite time $\tau$, such that in each realization
$x_t$ the work done is
$W=\int_0^\tau ds\, {\dot \lambda}_s\, \partial_\lambda
U(x_s,\lambda_s)$.
Due to the noise, the work $W$ is a fluctuating quantity. Yet, its
average $\mu_W[\lambda_t]=\langle W\rangle$ and standard deviation
$\sigma_W[\lambda_t]=\sqrt{\langle (W-\mu_W)^2\rangle}$ are uniquely
determined by the protocol $\lambda_t$.  Our goal is then, for fixed
protocol duration and end points
$(\tau,\lambda_{\rm i},\lambda_{\rm f})$, to characterize the
protocols that minimize either the average work, the standard
deviation, or their best compromise.

A general approach to the problem of optimizing a collection of
incompatible objectives that cannot be simultaneously optimal -- here,
$\mu_W$ and $\sigma_W$ -- is to utilize the notion of Pareto-optimal
solutions in order to classify all the possible optimal
protocols~\cite{Seoane2015b}. To this end, we will say that a protocol
$\lambda^1_t$ dominates another $\lambda^2_t$ if it performs better
for one of the objectives ({\it i.e.}\ leads to a smaller $\mu_W$ or
$\sigma_W$) and at least as well in the other objective. The
collection of Pareto-optimal protocols -- those that are not dominated
by any other protocol -- form the Pareto front and represent the set
of optimal solutions, for which one objective cannot be improved
without degrading the other. The Pareto front thus encodes the
possible trade-offs.  When plotted in the $\mu_W$-$\sigma_W$ plane, as
in Fig.~\ref{fig:illustration}(c), the Pareto-optimal solutions form a
boundary to the space of all feasible protocols.

A natural starting point for computing the Pareto front is to minimize
a single objective linear function~\cite{Seoane2015b}
\begin{equation}
\label{eq:Jalpha}{\mathcal J}_\alpha=\alpha \mu_W + (1-\alpha)\sigma_W,
\end{equation}
with $\alpha\in[0,1]$. As we vary $\alpha$ from $0\to 1$, we shift
from minimizing the standard deviation $\sigma_W$ to minimizing the
average work $\mu_W$~\cite{foot1}.  As illustrated in
Fig.~\ref{fig:illustration}(c), a protocol minimizing
${\mathcal J}_\alpha$ is always Pareto optimal. However, the converse
need not be true: the family of minima of ${\mathcal J}_\alpha$ maps
out the entire Pareto front only if the space of feasible protocols is
strictly convex~\cite{Seoane2015b}. For example, the green portion of
the front in Fig.~\ref{fig:illustration}(c) corresponds to a single
value of $\alpha$.  In the following, we will encounter both strictly
and not strictly convex fronts.

{\it Harmonic trap.---} As a first case study, we consider an
overdamped Brownian particle in a harmonic trap with potential
$U(x,k_t)=k_tx^2/2$, with controllable spring constant $k_t$.  We
choose this model for its 
tractability~\cite{Imparato2007,Schmiedl2008,Horowitz2009b} and 
its experimental
relevance~\cite{Blickle2011,Martinez2015,Martinez2016}.  The
particle's dynamics are given by the overdamped Langevin equation
\begin{equation}\label{eq:langevin}
\gamma {\dot x_t}=-k_tx_t+\sqrt{2\gamma k_{\rm B}T}\xi_t,
\end{equation}
where $\gamma$ is the viscosity, $T$ the temperature and $\xi_t$ is a
zero-mean, Gaussian white noise.  We optimize over protocols $k_t$ of
fixed duration $\tau$ with fixed initial and final values $k_{\rm i}$
and $k_{\rm f}$.  Choosing appropriate units, we can take
$k_BT=\gamma=1$ and express all results in terms of the ratio
$k_{\rm f}/k_{\rm i}$.

Under these constraints, we determine numerically the protocol
minimizing ${\mathcal J}_\alpha$ in Eq.~(\ref{eq:Jalpha}). This is
summarized as: (i) Exploiting the linearity of
Eq.~(\ref{eq:langevin}), we derive a closed set of ordinary
differential equations (ODEs), whose solution for a given protocol
outputs the mean work $\mu_W$ and standard deviation $\sigma_W$
(see~\cite{SI}). (ii) The ODEs are integrated by discretizing the
protocol into $N=100$ points with linear interpolations. We also
explicitly allow for discontinuities at $t=0$ and $t=\tau$, as these
are known to be generic for minimum-work 
protocols~\cite{Esposito2010, Aurell2011, Schmiedl2007,
  Gomez-Marin2008,Aurell2012}.  (iii) We then perform a stochastic
gradient descent to minimize ${\mathcal J}_\alpha$: At each step, a
small trial move $\delta k$ of one point of the protocol is proposed
and accepted if it decreases ${\mathcal J}_\alpha$. Remarkably, the
protocol space is found to be very smooth so that the optimization
procedure converges to a unique minimum independently of the initial
condition. This was checked for each $\alpha$ using 100 
random initial protocols, with each point drawn from a uniform
distribution on $[0,2k_{\rm f}]$.

\begin{figure}[tb]
\includegraphics[height=2.6cm,width=\linewidth]{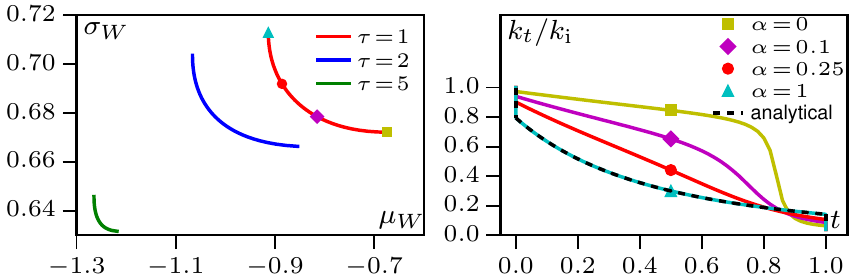}
\caption{{\bf Left:} Pareto front for the harmonic oscillator obtained
  by minimizing $\cJ_\alpha$ for three different protocol durations
  $\tau=1,2,5$. Squares on the $\tau=1$ curve indicate the
  position of the protocols shown on the right. {\bf Right:} The
  optimal protocols deform smoothly along the Pareto front as we vary
  $\alpha=0\to1$. The dashed black line indicates the exact analytical
  solution~\cite{Schmiedl2007}. Both plots are for an
  expansion $k_{\rm f}/k_{\rm i}=0.04$.}
\label{fig:osc}
\end{figure}

Repeating the process for different values of $\alpha$, we obtain the
family of solutions shown in Fig.~\ref{fig:osc} (left) for an
expansion with $k_{\rm f}/k_{\rm i}=0.04$ and protocol durations
$\tau=1,2,5$. We observe that varying $\alpha$ from $0$ to $1$, the
optimal protocols draw a continuous and convex line in the
$\mu_W$-$\sigma_W$ plane, which thus corresponds to the full family of
Pareto-optimal solutions. Correspondingly, the optimal protocols,
shown in Fig.~\ref{fig:osc}~(right) for $\tau=1$ deform smoothly along
the Pareto front. For $\alpha=1$, our algorithm recovers the
minimum-work protocol derived analytically in
Ref.~\cite{Schmiedl2007}. 
This minimum-work protocol smoothly decreases over the entire
interval (apart from discontinuous jumps at the edges), whereas the
minimum-fluctuation protocol stays relatively constant before dropping quickly.  By keeping the oscillator confined, the
small spread in position translates to a small spread in work values
during the final rapid expansion, despite costing
work~(cf.~\cite{Holubec2014}).

If we focus our attention on the optimal work protocols at $\alpha=1$,
we observe that the Pareto front's asymptote is vertical. This
indicates that to reach the optimal work protocol, one must sacrifice
a lot of fluctuations, relatively speaking. Put another way, there are
many near-optimal protocols with substantially less fluctuations,
complementing Ref.~\cite{Gingrich2016}, which found in a driven Ising
model that near-optimal protocols can be numerous.  Similarly, the
flat asymptote at $\alpha=0$ near the optimal-work-fluctuation
protocol indicates that a lot of dissipation is necessary to reduce
the fluctuations to a minimum.

\begin{figure}[tb]
\includegraphics[height=5cm,width=\linewidth]{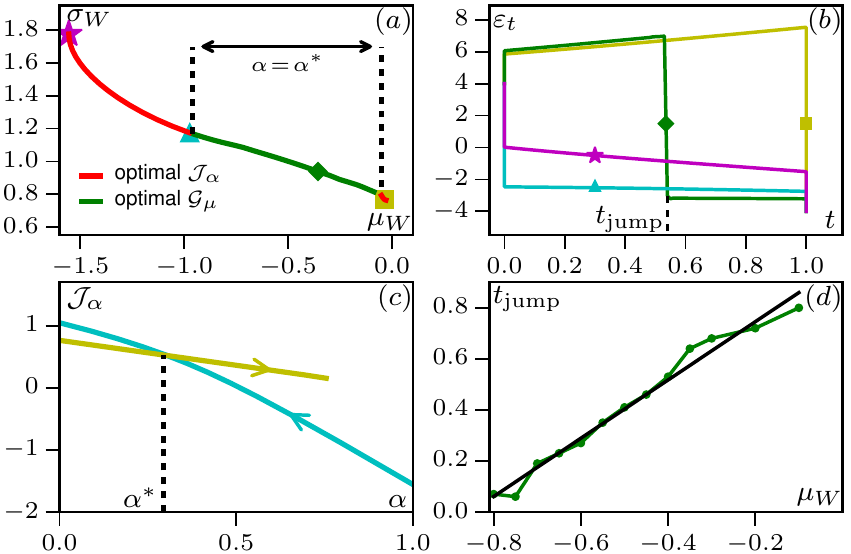}
\caption{{\bf (a):} Pareto front for the quantum dot, obtained by
  minimizing $\cJ_\alpha$ and $\cG_\mu$. Symbols indicate the position
  of the protocols shown in (b). {\bf (b):} Optimal work protocol
  (magenta), the two protocols at $\alpha=\alpha^*$ on the
  minimum-work and mimimum-work-fluctuation branches (blue and yellow)
  and a protocol in the coexistence region (green). {\bf (c):} Optimal
  ${\mathcal J}_\alpha$ obtained by ramping $\alpha$ up (yellow) or
  down (blue) without restarting from a random initial protocol, exchanging of global optimality at $\alpha^*$. {\bf (d):}
  Position of the discontinuity in phase coexistence protocols as a
  function of $\mu_W$ with a linear fit.
  Parameters: $\tau=1$, $\eps_{\rm i}=4$, $\eps_{\rm
    f}=-4$.}\label{fig:dot}
\end{figure}

{\it Two-level system.---} To allow for more complex behavior, we now
optimize a quantum dot~\cite{Esposito2010,Gopalkrishnan2016}.  We
model its dynamics as a Markov jump process with two discrete states,
empty or filled with one electron.  Jumps between states occur due to
the exchange of a particle with a reservoir. We denote by $\eps_t$ the
difference between the energy level of the dot and the chemical
potential of the reservoir. The system is fully characterized by the
probability $p_t$ to be filled, which evolves as (see
\cite{Esposito2010,SI})
\begin{equation}\label{eq:p} {\dot p}_t=-\omega p_t +
  \frac{\omega}{1+e^{\eps_t/k_{\rm B}T}},
\end{equation}
with bare rate constant $\omega$. We choose $\omega=k_{\rm B}T=1$,
fixing time and energy units.

Like the harmonic oscillator, the linearity of Eq.~(\ref{eq:p}) allows
us to construct a set of ODEs whose solution gives $\mu_W$ and
$\sigma_W$ for a protocol $\eps_t$ changing from $\eps_{\rm i}$ at
$t=0$ to $\eps_{\rm f}$ at $\tau$. The optimization procedure is
identical to that of the harmonic trap. We choose here a
representative set of parameters $\eps_{\rm i}=4$, $\eps_{\rm f}=-4$
and $\tau=1$. The protocols minimizing ${\mathcal J}_\alpha$ for
$\alpha\in [0,1]$ are shown in red in Fig.~\ref{fig:dot}(a). Strikingly,
as we vary $\alpha$ from $0$ to $1$, tracing the red line from bottom
right to top left, there is a discontinuous break (hopping over the
green segment), signaling a jump in protocol space (at
$\alpha^*\approx 0.305$ for our parameters): This corresponds
to a qualitative change in the optimal protocols pictured in
Fig.~\ref{fig:dot}(b); from minimum-fluctuation-like protocols with
$\eps_t$ increasing (apart from discontinuous jumps at the end
points), to minimum-work-like protocols with $\eps_t$ decreasing.  The
transition happens when these two different solutions that are locally
optimal in protocol space exchange global optimality.

The missing portion of the Pareto front can be accessed by optimizing a different function
\begin{equation}
  \label{eq:Gmu}
  \cG_\mu=\kappa(\mu_W-\mu_0)^2 +\sigma_W
\end{equation}
for fixed value of $\mu_0$. Taking large $\kappa=10$, the
protocol that minimizes $\cG_\mu$ has an average work very close to
the fixed value $\mu_W\approx \mu_0$ and minimum standard
deviation. It is thus a good approximation of the point on the
Pareto front at $\mu_0$. (An alternative method
imposing a hard inequality constraint can be found in
Ref.~\cite{zitzler1999evolutionary}.) Varying $\mu_0$ then yields the
green portion of Fig.~\ref{fig:dot}(a), thereby completing the
front. The resulting protocols, as shown in Fig.~\ref{fig:dot}(b),
exhibit a sharp jump in the middle. Numerically, we find that in this
part of the phase diagram our stochastic gradient descent can get
trapped in local minima corresponding to different positions of the
jump. To find the global minimum of $\cG_\mu$, we thus performed many
runs ($>500$) with different initial conditions to sample all
local minima. For more precision, we also adapted our code to replace
sharp gradients by exact discontinuities.

\begin{figure}[tb]
\includegraphics[height=2.6cm,width=\linewidth]{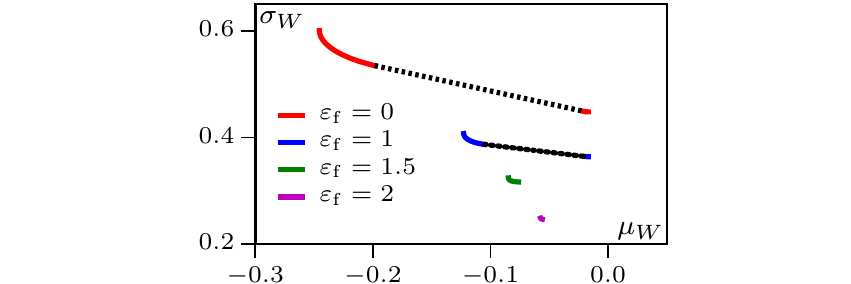}
\caption{Pareto front of the quantum dot for
  varying $\eps_{\rm f}$ at fixed $\eps_{\rm i}=4$ and $\tau=1$. For
  $\eps_{\rm f}>1$, the optimal-work-fluctuation solution family 
  disappears and the Pareto front becomes strictly convex.  Dotted
  black lines are guides to the eye denoting the coexistence
  regions.}
\label{fig:dot-epsf}
\end{figure}

Putting everything together, the picture is similar to that of a
first-order liquid-gas transition~\cite{Seoane2015b}. The parameter
$\alpha$ in $\cJ_\alpha$ plays the role of an intensive parameter (say
the pressure), whereas $\cJ_\alpha$ is analogous to the free
  energy: There is a finite jump to a
different protocol at a critical $\alpha^*$ where the two solutions
exchange global optimality.  As in a liquid-gas transition,
these ``homogeneous'' solutions remain metastable beyond
$\alpha^*$. Although these suboptimal solutions are not part of the
Pareto front, they can be accessed by slowly ramping $\alpha$ while
minimizing $\cJ_\alpha$ without restarting from a random initial
protocol, in the same way as hysteresis loops are observed by
ramping fluid pressure. The transition point $\alpha^*$
corresponds to the exchange of global optimality, as shown in
Fig.~\ref{fig:dot}(c).

Minimizing $\cG_\mu$ is then akin to switching to a constant-volume
(canonical) ensemble. This allows us to observe the analog of phase
coexistence {\it inside} the protocols: One observes a family of
protocols (all at $\alpha=\alpha^*$) that comprise two parts --
decreasing minimum-work-like and increasing minimum-fluctuation-like
-- linked together by a discontinuity. As shown in
Fig.~\ref{fig:dot}(d), the proportions of each ``phase'' vary linearly
along the front (up to numerical uncertainty), similar to what the
lever rule predicts for a liquid-gas transition.

\begin{figure}
\includegraphics[height=2.6cm,width=\linewidth]{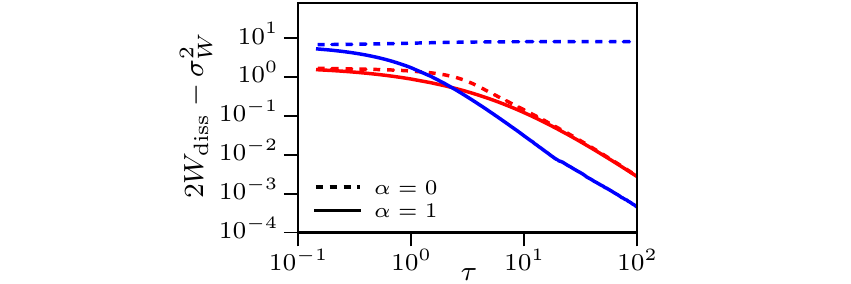}
\caption{Approach to the linear-perturbation regime for optimal work and
  work-fluctuation protocols as $\tau$ increases. Harmonic oscillator
  (red lines) with $k_{\rm f}/k_{\rm i}=0.04$ and quantum dot (blue
  lines) with $\eps_{\rm i}=4$, $\eps_{\rm f}=-4$. }
\label{fig:wdiss}
\end{figure}

The two ``homogeneous'' solutions correspond to two distinct
strategies: (i) The optimal work is achieved by monotonically
decreasing the energy level while (ii) the minimum deviation is
achieved by first increasing the energy to confine the system into a
single discrete state with almost no spread in state space prior to a
rapid fluctuationless switch.  Physically, in case (i) the dot is
partially filled and the protocol tries to keep the distribution as
much like the equilibrium distribution as possible.  In case (ii), the
dot is mostly empty and thus ends with a distribution very different
from the final equilibrium.  The trapping of the distribution, made
possible by the system's discreteness, is at the origin of the
transition. This is confirmed by Fig.~\ref{fig:dot-epsf}, which shows
the Pareto fronts for different  $\eps_{\rm f}$ at fixed
$\eps_{\rm i}=4$. For larger $\eps_{\rm f}>1$, the initial and final
energy levels are not separated enough to make the compressed
distribution very different from the final equilibrium.  Consequently,
the optimal protocols approach the linear regime where work
fluctuations are constrained to be equal to the average
work~\cite{Hermans1991,Then2008}, forcing the optimal-work-fluctuation
branch to disappear.

{\it Quasistatic limit.---}
This disappearance of the transition suggests a similar phenomenon would occur for the linear regime reached for long times~\cite{Rahav2008,Horowitz2009,Zulkowski2012,Sivak2012,Bradner2015,Proesmans2016,Rotskoff2017}.
Thus, we would expect all optimal protocols to collapse onto a quasistatic one that remains nearly in
equilibrium at every point in time.
The Second Law, however, only guarantees that minimum-work protocols 
become quasistatic, whereas this need not be true for protocols optimizing
a different quantity. Indeed, we show here that the protocols
minimizing work fluctuations for the quantum dot never become
quasistatic.

Close to the quasistatic limit, linear response predicts that
$2W_{\rm diss}=\sigma_W^2$, with $W_{\rm diss}=\mu_W-\Delta F$ the
dissipated work and $\Delta F$ the free energy difference between the
initial and final equilibrium~\cite{Hermans1991,Speck2004}. For
the harmonic oscillator, Fig.~\ref{fig:wdiss} shows that as $\tau$
increases the linear response regime is approached by both the optimal-work and optimal-work-fluctuation protocols. On the contrary, for the
quantum dot, only the optimal-work protocol approaches the linear
 regime. Even in the infinite time limit, the optimal-work
fluctuation-protocol dissipates a finite amount and thus
remains nonquasistatic.

\begin{figure}
\includegraphics[height=2.6cm,width=\linewidth]{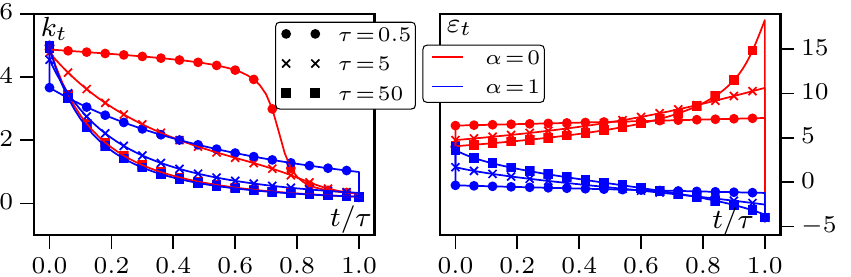}
\caption{Optimal work ($\alpha=1$) and work-fluctuation ($\alpha=0$)
  protocols of varying duration $\tau$. {\bf Left:} Harmonic
  oscillator with $k_{\rm f}/k_{\rm i}=0.04$. The two protocols become identical
  in the large duration limit. {\bf Right:} Quantum dot with
  $\eps_{\rm i}=4$, $\eps_{\rm f}=-4$. The two protocols remain different for
  large durations.}
\label{fig:large-tau}
\end{figure}

The different limits are best understood by looking at how the optimal
protocols change as $\tau$ increases, as shown in
Fig.~\ref{fig:large-tau}. For the harmonic oscillator, the whole
Pareto front contracts to a point in protocol space as the two
extremities for $\alpha=0$ and $\alpha=1$ converge to the same
protocol. On the contrary, the protocols corresponding to minimal mean
work and work standard deviation remain different as $\tau\to\infty$
for the quantum dot. The structure of the phase diagram of
Fig.~\ref{fig:dot} is preserved upon increasing $\tau$ so that there
always exists two ``phases''. Only the family containing the optimal
work protocol becomes quasistatic for large $\tau$ while the optimal
fluctuation protocols retain instantaneous jumps and are therefore not
quasistatic. Thus, studies of optimal work fluctuation protocols
within linear irreversible thermodynamics cannot access the optimal
solution.

To summarize, we have shown that for the harmonic oscillator, the
trade-off between work and work fluctuations is captured by a smooth
family of protocols. However, this behavior is not generic. For a
quantum dot, approximating a double-well potential, optimal work and
work-fluctuation protocols belong to qualitatively different
``phases''. The trade-offs between the two, captured by the Pareto
front, have the structure of a first-order phase transition with
phase-coexistence protocols interpolating between the two phases.
Such a phase transition may be a common feature of optimization
problems: they occur in optimal complex
networks~\cite{Seoane2015,Seoane2016}, statistical inference
\cite{zdeborova2016statistical}; and similar phenomena were observed
in a quantum control problem with varying constraints~\cite{Bukov2017}
and in the utilization of memory in an information
engine~\cite{hartich2014stochastic,bechhoefer2015hidden,lathouwers2017memory}.
Finally, we observed that the minimum work fluctuation and the phase
coexistence protocols do not become quasistatic even for very long
protocols and are thus not accessible by a linear theory.

\begin{acknowledgments}
  We are especially grateful to Todd Gingrich and Luis
Seoane for their advice and suggestions and acknowledge the Gordon and
Betty Moore Foundation for supporting us as Physics of Living
Systems Fellows through Grant GBMF4513.
\end{acknowledgments}

\bibliography{OptWorkRef}

\begin{thebibliography}{63}
\bibitem [{foo()}]{foot1}%
  \BibitemOpen
  \bibfield  {title} {\enquote {\bibinfo {title} {We choose to use the standard
  deviation $\sigma_w$ so that all quantities in ${\mathcal j}_\alpha$ have the
  same dimension although this is not necessary and using the variance
  $\sigma_w^2$ instead would yield the same results.}}\ }\href@noop {} {\
  }\BibitemShut {NoStop}%
\end{thebibliography}

\end{document}